\documentclass[11pt]{article}
\usepackage[usenames,dvipsnames,svgnames,table]{xcolor} 

\usepackage[pdfstartview=XYZ,
bookmarks=true,
colorlinks=true,
linkcolor=blue,
urlcolor=blue,
citecolor=blue,
pdftex,
bookmarks=true,
linktocpage=true, 
hyperindex=true
]{hyperref}
\usepackage{orcidlink}
\usepackage{hyperref}
\usepackage{siunitx}
\usepackage{listings}
\usepackage{courier}
\usepackage{color}
\definecolor{dkgreen}{rgb}{0,0.6,0}
\definecolor{gray}{rgb}{0.5,0.5,0.5}
\definecolor{mauve}{rgb}{0.58,0,0.82}
\usepackage{algorithm}
\usepackage{algorithmic}
\floatname{algorithm}{Algorithm}

\usepackage{longtable}
\usepackage{hhline}
\usepackage{float}
\newfloat{algorithm}{tb}{lop}

\usepackage{enumitem} 
\usepackage{rotfloat} 
\usepackage{graphicx}
\usepackage{epsfig}
\usepackage{amssymb, amsmath, amsthm}
\usepackage{verbatim}
\usepackage{natbib}
\usepackage{authblk}
\usepackage{multirow}
\usepackage{subcaption} 
\usepackage{array}
\newcolumntype{L}{>{\centering\arraybackslash}m{0.1\linewidth}}

\newtheorem{theorem}{Theorem}

\newtheorem{proposition}[theorem]{Proposition}
\newtheorem{corollary}[theorem]{Corollary}

\DeclareMathOperator*{\argmax}{argmax}

\newcommand{\calU}{\mathcal{U}}

\newcommand{\bbR}{\mathbb{R}}
\newcommand{\bbS}{\mathbb{S}}

\newcommand{\bfone}{\mathbf{1}}

\usepackage{tcolorbox}
\newtcbox{\mybox}[1][]{%
	nobeforeafter,
	colframe=gray,
	boxrule=0.5pt,
	minipage,
	width=0.95\linewidth, 
	valign=center,
	height=1.8cm,               
	arc=3pt,
	boxsep=0pt,
	#1
}
\newtcbox{\mybigbox}[1][]{%
	nobeforeafter,
	colframe=gray,
	boxrule=0.5pt,
	minipage,
	width=0.95\linewidth, 
	valign=center,
	height=3cm,               
	arc=3pt,
	boxsep=0pt,
	#1
}

\begin{document}
	
\title{PCA, SVD, and Centering of Data}
\author[1]{Donggun Kim}
\author[2]{Kisung You\,\orcidlink{0000-0002-8584-459X}}
\affil[1]{Independent Researcher}
\affil[2]{Department of Mathematics, Baruch College}
\date{}

\maketitle
\begin{abstract}

The research detailed in this paper scrutinizes Principal Component Analysis (PCA), a seminal method employed in statistics and machine learning for the purpose of reducing data dimensionality. Singular Value Decomposition (SVD) is often employed as the primary means for computing PCA, a process that indispensably includes the step of centering - the subtraction of the mean location from the data set. In our study, we delve into a detailed exploration of the influence of this critical yet often ignored or downplayed data centering step. Our research meticulously investigates the conditions under which two PCA embeddings, one derived from SVD with centering and the other without, can be viewed as aligned. As part of this exploration, we analyze the relationship between the first singular vector and the mean direction, subsequently linking this observation to the congruity between two SVDs of centered and uncentered matrices. Furthermore, we explore the potential implications arising from the absence of centering in the context of performing PCA via SVD from a spectral analysis standpoint. Our investigation emphasizes the importance of a comprehensive understanding and acknowledgment of the subtleties involved in the computation of PCA. As such, we believe this paper offers a crucial contribution to the nuanced understanding of this foundational statistical method and stands as a valuable addition to the academic literature in the field of statistics.
\end{abstract}


\pagebreak 
\section{Introduction}\label{sec:intro}

Principal component analysis \citep{pearson_1901_LIIILinesPlanes}, abbreviated as PCA, is one of the most fundamental methods in statistics and related disciplines. Its main usage includes reducing the dimensionality of data beyond human perceptual boundaries for visualization and downstream analysis \citep{joliffe_2002_PrincipalComponentAnalysis}. Owing to its simple yet effective nature in representing high-dimensional data in a more compact form, PCA has been used across multiple areas of science, with numerous implementations in many programming languages.

Suppose we are given a data matrix $X \in \bbR^{n\times p}$, consisting of $n$ observations, each represented as a $p$-dimensional vector. The derivation of PCA directly leads to a sequential algorithm that applies eigendecomposition to an (empirical) covariance matrix derived from the data. While straightforward, this method has an inherent drawback: when the data dimensionality $p$ is high, it requires an excessive amount of computational resources. A popular alternative is to apply the Singular Value Decomposition (SVD) to the data matrix after subtracting the mean location.

In the latter statement, one might notice that SVD-based computation requires subtraction of the mean location, often termed mean centering. While this is a necessary requirement to guarantee exact equivalence to the direct computational scheme, this step is often overlooked or not even mentioned, as is often observed in online resources or from students with whom the authors have interacted. There are also heuristics, such as a folklore theorem, that suggest employing an extra base from the SVD, based on the argument that the first right singular vector accounts for the mean direction. This claim may be partially correct under restricted circumstances.

In this paper, we rigorously investigate the aforementioned claims and study when and how two PCA embeddings from the SVD, with and without centering, can be considered \textit{aligned}. We begin by examining the relationship between the first singular vector and the mean direction, and connecting this observation to the equivalence of SVDs of centered and uncentered matrices. Finally, we study this topic from a spectral point of view, which provides, at least, a partial answer to the question of what happens if PCA is performed via SVD without centering the data matrix. This work is an extended version of \cite{kimPeriodicSignalAnalysis2015}.

\section{Background}\label{sec:twoschemes}

We begin by formally introducing two computational schemes. Let $X\in\bbR^{n\times p}$ be a data matrix of interest, consisting of $n$ observations $\lbrace x_1, \ldots, x_n\rbrace $ in $\bbR^p$. In other words, each row corresponds to an observation. The goal of PCA is to project this high-dimensional data from $\bbR^p$ onto a lower-dimensional space $\bbR^k$ where $k<p$, where a low-dimensional embedding will be denoted as $Y \in \bbR^{n\times k}$. For the sake of simplicity, we will assume that $p < n$ throughout the paper.

\subsection{Two schemes}

Now, we turn our attention to detailing the two schemes mentioned earlier. The first scheme involves the use of eigendecomposition of an empirical covariance matrix. This approach is likely the most straightforward way to both understand the nature of PCA and implement the method using modern computing platforms. Note that $\bfone_n = [1,\ldots, 1] \in \mathbb{R}^n$ represents a vector of length $n$ whose entries are all 1's, and $\bar{x} = \sum_{i=1}^n x_i / n$ denotes a mean vector.

\vspace{0.25cm}
\begin{tcolorbox}
	\begin{center}
		\textbf{Scheme 1} : PCA by eigendecomposition of an empirical covariance matrix
	\end{center}
	\vspace{-.15cm}
	\hrule
	\begin{description}
		\item[Step 1.] Compute an empirical covariance matrix $\hat{\Sigma} \in \bbR^{p\times p}$,
		\begin{equation*}
			\hat{\Sigma} = \frac{1}{n-1}(X - \bfone_n \bar{x}^\top)^\top(X - \bfone_n \bar{x}^\top)
		\end{equation*}
		\item[Step 2.] Apply eigendecomposition to $\hat{\Sigma}$,
		\begin{equation*}
			\hat{\Sigma} V = V \Lambda \quad \longrightarrow \quad  \hat{\Sigma} v_j = \lambda_j v_j\text{ for }j=1,\ldots, p.
		\end{equation*}
		The eigenvalues are ordered, i.e., $\lambda_1 \geq \cdots \geq \lambda_p$ and we assume the same ordering for eigenvectors $v_1, \ldots, v_p$. 
		\item[Step 3.] Assemble the projection matrix $V_{1:k}$ by taking the first $k$ eigenvectors,
		\begin{equation*}
			V_{1:k} = [v_1, \ldots, v_k] \in \bbR^{p\times k}.
		\end{equation*}
		\item[Step 4.] Compute the low-dimensional embedding $Y = X V_{1:k} \in \mathbb{R}^{n\times k}$. 
	\end{description}
\end{tcolorbox}

A primary drawback of \textbf{Scheme 1} is its unfeasibility when the dimensionality is high - that is, when $p$ is large. Computing the empirical covariance matrix becomes impractical. In the \textsf{R} programming environment \citep{rcoreteam_2022_LanguageEnvironmentStatistical}, for instance, a dimensionality of $p=10^4$ results in a covariance matrix that consumes approximately \qty{190}{\mega\byte} of memory. If $p$ is increased to $5 \times 10^4$, memory usage escalates to \qty{18.6}{\giga\byte} in the standard dense matrix format. These constraints underscore the necessity for an efficient and feasible alternative, prompting the use of singular value decomposition. The algorithmic details of this alternative approach are described below.

\vspace{0.25cm}
\begin{tcolorbox}
	\begin{center}
		\textbf{Scheme 2} : PCA by SVD
	\end{center}
	\vspace{-.15cm}
	\hrule
	\begin{description}
		\item[Step 1.] Center the data matrix,
		\begin{equation*}
			\bar{X} = X - \bfone_n \bar{x}^\top.
		\end{equation*}
		\item[Step 2.] Apply singular value decomposition to $\bar{X}$,
		\begin{equation*}
			\bar{X} = U \Sigma V^\top.
		\end{equation*}
		Note that left and right singular vectors $u_i, v_j$ are ordered according to the descending order of singular values $\sigma_1 \geq \ldots \geq  \sigma_p$, respectively.
		\item[Step 3.] Assemble the projection matrix $V_{1:k}$ by taking the first $k$ right singular vectors,
		\begin{equation*}
			V_{1:k} = [v_1, \ldots, v_k] \in \bbR^{p\times k}.
		\end{equation*}
		\item[Step 4.] Compute the low-dimensional embedding $Y = X V_{1:k} \in \mathbb{R}^{n\times k}$. 
	\end{description}
\end{tcolorbox}

When the data is centered -- $\|\bar{x}\| = 0$ -- or centering is applied directly to the data, \textbf{Scheme 2} avoids the need to create and store an empirical covariance matrix $\hat{\Sigma}$ in memory.

If the data is centered, it is straightforward to see that the two schemes are equivalent. Suppose the given data $X$ is centered, i.e., $\bar{x} = 0$, with an SVD of $X = U\Sigma V^\top$. This implies $X^\top X = V \Sigma^\top U^\top U \Sigma V^\top = V \Sigma^2 V^\top$. Given that the empirical covariance of centered $X$ is $X^\top X / (n-1)$, it becomes apparent that the eigenvectors of an empirical covariance matrix are equivalent to the right singular vectors of the original data matrix.

\subsection{No centering, No equivalence}\label{subsec:noequivalence}

The SVD approach is \textit{theoretically valid} if and only if the data is centered. However, some textbooks and tutorials may gloss over this prerequisite, instead prefacing their explanations with a statement such as``assume the data is centered.'' This approach can potentially lead to confusion for those without sufficient experience or who are not paying careful attention.
For instance, in the \textsf{R} programming environment, \textbf{Scheme 1} can be implemented in just four lines of code\footnote{In fact, a single line of code is all you need: {\ttfamily Y = X\%*\%eigen(cov(X))\$vectors[,1:k]}.}.

\begin{lstlisting}[title={Implementation of \textbf{Scheme 1} using \textsf{R}.}, label=code:scheme1]
	Sig <- cov(X)          # step 1 : covariance computation 
	eig <- eigen(Sig)      # step 2 : eigendecomposition
	V <- eig$vectors[,1:k] # step 3 : assemble the projection matrix
	Y <- X%*%V             # step 4 : projection
\end{lstlisting}

Observe that the computation of covariance is managed by the {\ttfamily cov()} function, which automatically performs mean centering. This is a common feature in many other programming languages. However, in the implementation of \textbf{Scheme 2}, the centering of the data must be executed explicitly, necessitating additional care. Without this step, the two projection matrices cannot be equivalent (in general).

\begin{figure}[ht]
	\centering
\begin{subfigure}{.33\textwidth}
	\centering
	\includegraphics[width=.99\linewidth]{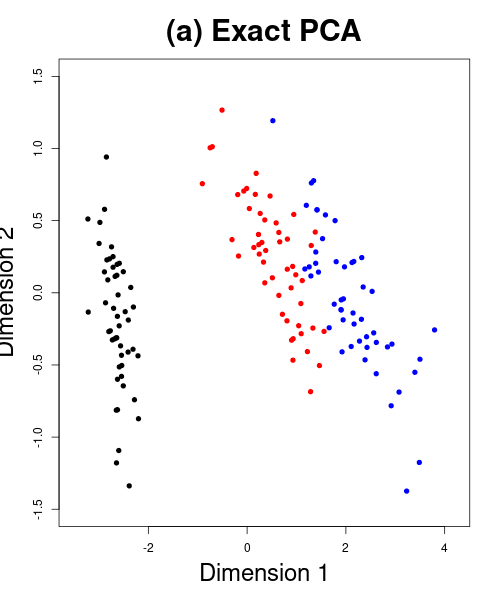}
\end{subfigure}%
\begin{subfigure}{.33\textwidth}
	\centering
	\includegraphics[width=.99\linewidth]{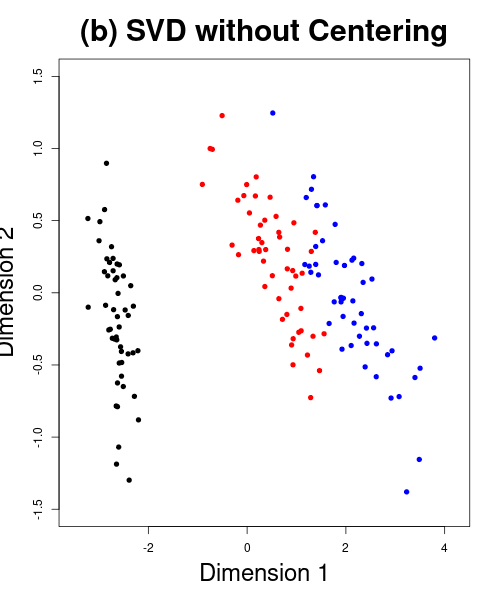}
\end{subfigure}
\begin{subfigure}{.33\textwidth}
	\centering
	\includegraphics[width=.99\linewidth]{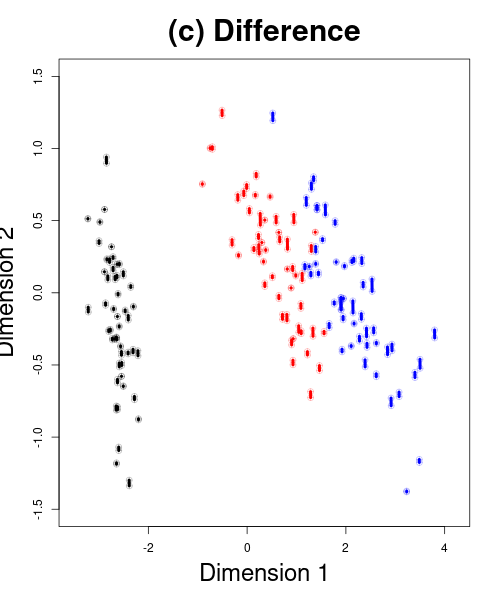}
\end{subfigure}
\caption{Visualization of the \textit{iris} dataset demonstration. Two embeddings obtained by \textbf{Scheme 1} and \textbf{Scheme 2} are shown in (a) and (b), respectively. Their discrepancy after normalization is visualized in (c). Three distinct colors indicate categories of the flowers.}
\label{fig:iris}
\end{figure}

A reasonable question arises: what would happen to the SVD-based algorithm if the centering step were omitted? Theoretically, this would lead to an incorrect realization of PCA. We illustrate this discrepancy with a simple example using the \textit{iris} dataset \citep{fisher_1936_USEMULTIPLEMEASUREMENTS, anderson_1936_SpeciesProblemIris}. This multivariate dataset consists of four-dimensional measurements taken from 150 samples of \textit{Iris} flowers, each belonging to one of three classes. After verifying that the mean vector of the data is not zero, we computed low-dimensional embeddings using \textbf{Scheme 1} and \textbf{Scheme 2 without centering}. These embeddings were then normalized - mean-centered to have zero means - and aligned using the orthogonal Procrustes matching \citep{dryden_1998_StatisticalShapeAnalysis} method, facilitating a direct comparison of their shapes. These steps are all illustrated in Figure \ref{fig:iris}.

It is surprising to observe that the two normalized embeddings are indeed very similar to each other. However, as shown in Figure \ref{fig:iris}(c), a clear discrepancy between the two does exist. While a small number of points appear visually identical, the majority of embedded coordinates are displaced to varying degrees. These observations lead us to question the relationship between the two schemes, prompting us to investigate the conditions under which the SVD-based approach without centering approximates the exact PCA.

\section{Main}\label{sec:main}

The observations from the previous section prompt us to investigate how the embeddings from two schemes are related when the data is not centered or when the centering step is overlooked. For the remainder of this paper, we will assume that the data $X$ has a non-zero mean and will denote the SVDs of the original data matrix $X$ and its centered counterpart $\bar{X}$ as follows:

\begin{equation*}
	X = U\Sigma V^\top = \sum_{i=1}^{p} \sigma_i u_i v_i^\top\quad\text{and}\quad \bar{X} = \bar{U} \bar{\Sigma} \bar{V}^\top = \sum_{j=1}^p \bar{\sigma}_j \bar{u}_j \bar{v}_j^\top.
\end{equation*}

As our primary interest is in the right singular vectors, we will use `singular vectors' to refer to right singular vectors, using the terms `left' and `right' only when necessary for clarification. Similarly, we will use phrases such as `centered singular values' and `vectors' to denote those obtained from the SVD of $\bar{X}$.

\subsection{Mean direction and first singular vector}

One of the widespread folk theorems in the literature of PCA is that the first singular vector $v_1$ is almost equivalent to the mean direction $\bar{x}$ when $\| \bar{x} \|$ is large. We take a bit different approach in that an identity of $v_1$ and $\bar{v}_1$ in connection to $\bar{x}$ is first established as follows. 

\vspace{0.25cm}
\begin{tcolorbox}[colback=GreenYellow]
\begin{proposition}\label{theory:parallel}
	If the first centered singular vector $\bar{v}_1$ is parallel to $\bar{x}$ when $\|\bar{x}\|\neq 0$, then two singular vectors are equal, i.e., $v_1 = \bar{v}_1 = \bar{x}/\|\bar{x}\|.$
\end{proposition}
\end{tcolorbox}
\vspace{0.25cm}

\begin{figure}[ht]
	\centering
	\includegraphics[width=.5\linewidth]{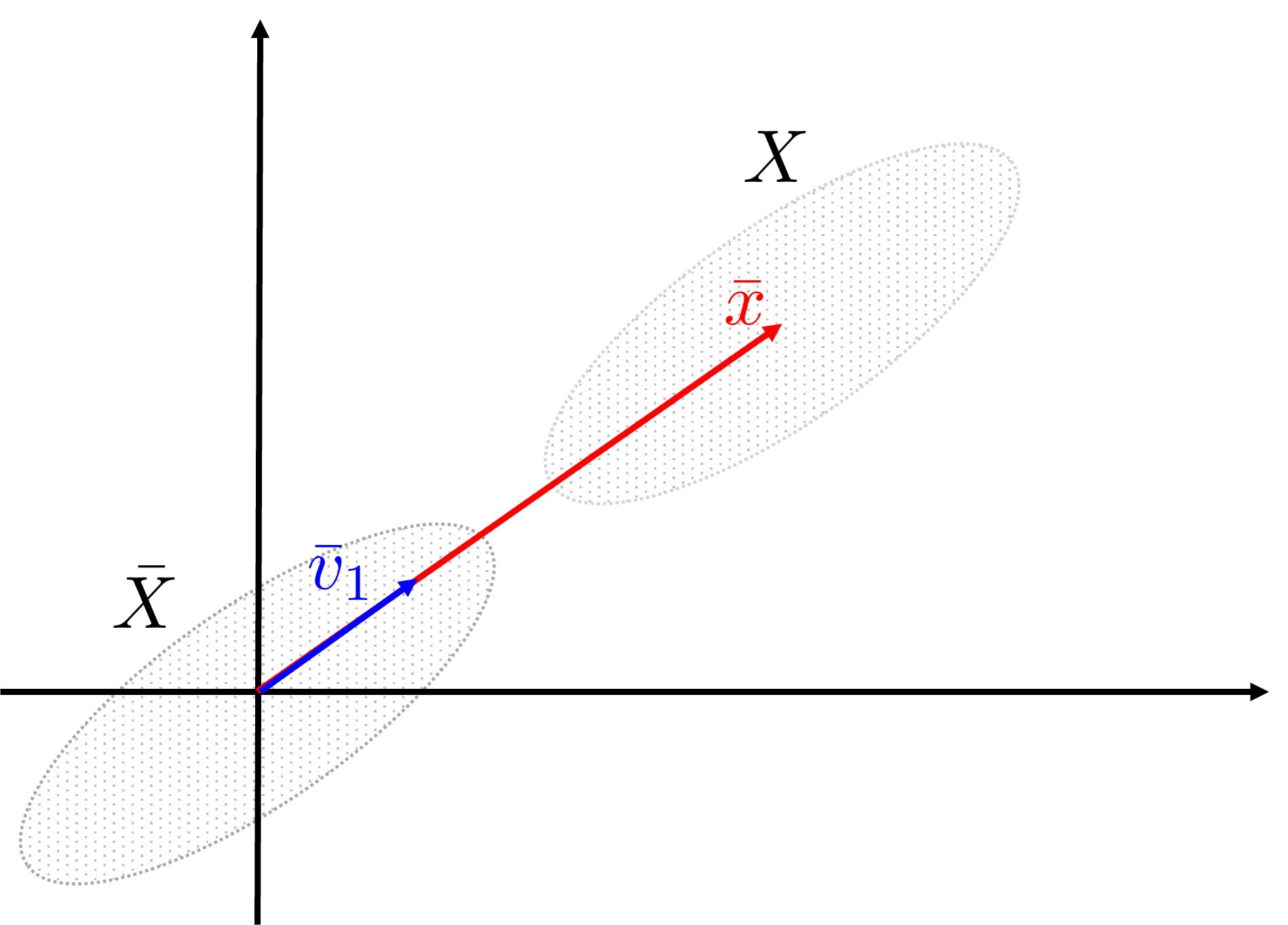}
	\caption{An exemplary visualization when $\bar{x}$ and $\bar{v}_1$ are parallel.}
	\label{fig:proposition1}
\end{figure}

We make an intuitive observation for the implication of Proposition \ref{theory:parallel} as follows as shown in Figure \ref{fig:proposition1}. Suppose $\bar{v}_1$ and $\bar{x}$ are parallel. Then, the first right singular vector $v$ also lies on the subspace spanned by $\bar{v}_1$ or equivalently that of $\bar{x}$. 

An immediate by-product of Proposition \ref{theory:parallel} is that two decompositions of $X$ and $\bar{X}$ are ``very close'' when the parallel condition is met.

\vspace{0.25cm}
\begin{tcolorbox}[colback=GreenYellow]
\begin{corollary}\label{theory:corollary2}
	Under the assumption of Proposition \ref{theory:parallel}, two SVDs of $X$ and $\bar{X}$ are identical except for the first singular value and the corresponding left singular vector. 
\end{corollary}
\end{tcolorbox}
\vspace{0.25cm}

Corollary \ref{theory:corollary2} provides a sufficient condition for two SVDs of $X$ and $\bar{X}$ to be equivalent that $\bar{v}_1$ and $\bar{x}$ should be parallel. This implies that the first principal component can be retrieved by SVD on a non-centered matrix only if the first eigenvector of $X^\top X$ is parallel to the mean vector.

\subsection{Mean direction and proximity to the first singular direction}

The previous section examines the relationship between $\bar{v}_1$ and $\bar{x}$ with focus on how it engenders equivalent SVDs of $X$ and $\bar{X}$. It is natural for a practitioner to ask whether my data abides by the reasoning. Unfortunately, this is type of knowledge one cannot acquire prior to actually applying SVD onto the data. In this section, we study when $v_1$ and $\bar{x}$ are close via providing a lower bound for the mean vector to be equivalent to the first eigenvector of $X^\top X$.

\vspace{0.25cm}
\begin{tcolorbox}[colback=GreenYellow]
	\begin{proposition}\label{theory:shape1}
		Let $z$ be an arbitrary unit vector in $\bbR^p$ and $z_0 = \bar{x}/\|\bar{x}\|$ be the scaled mean vector of unit norm. If the following inequality holds  with respect to the mean vector holds
		\begin{equation}\label{eq:lower_bound}
			\|\bar{x}\|^2 \geq \frac{\bar{\sigma}_{\max}^2-\bar{\sigma}_{\min}^2}{n(1-(\langle z_0, z\rangle)^2)},
		\end{equation}
		where $\bar{\sigma}_{\max}^2$ and $\bar{\sigma}_{\min}^2$ are the largest and smallest eigenvalues of $\bar{X}^\top \bar{X}$ respectively, then $z_0$ is indeed the largest eigenvector of $X^\top X$. 
	\end{proposition}
\end{tcolorbox}
\vspace{0.25cm}

Proposition \ref{theory:shape1} tells about under what circumstances a mean direction corresponds to the largest eigenvector of $X^\top X$ or the first right singular vector of a non-centered data matrix $X$ interchangeably. The inequality \eqref{eq:lower_bound} is more likely to hold when the numerator is close to 0 or the denominator goes to $\infty$. The former pertains to the \emph{eccentricity of the data} as $\bar{\sigma}_{\max}$ and $\bar{\sigma}_{\min}$ represent magnitudes of variation onto directions of the first and last principal components, respectively. A more interesting observation is that as $n \rightarrow \infty$, the inequality is more likely to hold without any distributional assumptions.

\vspace{0.25cm}
\begin{tcolorbox}[colback=GreenYellow]
\begin{corollary}\label{theory:shape2}
	For any $\epsilon \in (0,1)$, define $\calU_{\epsilon} = \lbrace u \in \bbS^{p-1}~|~ |u^\top z_0| \geq 1-\epsilon\rbrace $ and denote the first right singular vector of $X$ as $v_1$. If the following inequality holds
	\begin{equation*}
		\|\bar{x}\|^2 \geq \frac{\bar{\sigma}_{\max}^2 - \bar{\sigma}_{\min}^2}{n\cdot \left(1 - (1-\epsilon)^2\right)},
	\end{equation*}
	then $v_1 \in \calU_{\epsilon}$. 
\end{corollary}
\end{tcolorbox}
\vspace{0.25cm}

Corollary \ref{theory:shape2} provides heuristics to speculate when the first singular vector can be close to the mean direction, or vice versa. This suggests that proximity between first singular vector $v_1$ and the mean direction $\bar{x}$ is guaranteed under non-asymptotic regime given the norm condition.

\subsection{Spectral discrepancy between two SVDs}

When the first centered singular vector is parallel to the mean direction, it was witnessed that two SVDs of centered and uncentered data matrix are almost identical. The extent of equivalence includes the right singular vectors, which characterizes a basis for low-dimensional projection. Under such circumstances, the next question pertains to preservation of variances after projection. In the literature of PCA, eigenvalues of an empirical covariance matrix are quantities that account for variance on the 1-dimensional projection onto principal components. We note that there is an interesting phenomenon with respect to eigenvalues of $X^\top X$ and $\bar{X}^\top \bar{X}$, which are covariance matrices of uncentered and centered data respectively. Here we write with a simple fact that eigenvalues of $X^\top X$ are equivalent to squared singular values of $X$.

\vspace{0.25cm}
\begin{tcolorbox}[colback=GreenYellow]
\begin{proposition}\label{prop:3}
	If the first centered singular vector $\bar{v}_1$ is parallel to $\bar{x}$, eigenvalues of $X^\top X$ and $\bar{X}^\top \bar{X}$ have interlacing properties, i.e., 
	\begin{equation*}
		\bar{\sigma}_p^2 < \sigma_p^2 < \cdots < \bar{\sigma}_1^2 < \sigma_1^2 < \bar{\sigma}_1^2 + n \|\bar{x}\|^2,
	\end{equation*}
	where $\lbrace\sigma_i\rbrace$ and $\lbrace \bar{\sigma}_i\rbrace$ for $i=1,\ldots,p$ are singular values of $X$ and $\bar{X}$, respectively. 
\end{proposition}
\end{tcolorbox}
\vspace{0.25cm}

When the centering assumption is not abided, it says that the first singular value $\sigma_1$ absorbs the mean effect. This explains discrepancy of projected variance onto the first principal component with respect to the mean $\bar{x}$ and cardinality $n$ of the data. The rest of singular values are placed between the principal values of adjacent indices, which engenders the name of the \textit{interlacing} property.

Lastly, we provide a simple yet intriguing result on discrepancy between the two matrices in terms of their spectral norms. Recalling that   $X = \bar{X} + \bfone_n \bar{x}^\top$, denote the partial sum of the first $k$ squared singular values of a matrix $A$ as $\|A\|_{2,k}^2$.

\vspace{0.25cm}
\begin{tcolorbox}[colback=GreenYellow]
\begin{theorem}\label{theory:partialSum}
	Assume that the first singular vector $v_1$ equals the mean direction, whose squared norm $\|\bar{x}\|^2$ is \textit{large} in the sense of Corollary \ref{theory:shape2}. Then, the following statement holds for $k=1,\ldots,p-1$,
	\begin{equation*}
		-\bar{\sigma}_1^2 + \bar{\sigma}_{k+1}^2 < \|X\|_{2,k+1}^2 - 
		(\|\bar{X}\|_{2,k}^2 + n \|\bar{x}\|^2) < \bar{\sigma}_1^2.
	\end{equation*}
\end{theorem}
\end{tcolorbox}
\vspace{0.25cm}

Theorem \ref{theory:partialSum} provides a new perspective on the equivalence between the SVDs of $X$ and $\bar{X}$. It suggests that the difference of the partial sums of squared singular values for $X$ and $\bar{X}$ plus the mean effect is minimal. Note that the partial sums have different indices for $X$ and $\bar{X}$. This provides an approximate interpretation of a common convention or folkloric belief: that it suffices to use the second to the $(k+1)$-th right singular vectors of a data matrix without centering to project onto the $k$-dimensional Euclidean space.

An intriguing corollary of Theorem \ref{theory:partialSum} is the case when a given data matrix is of low rank. In linear algebra, it is a well-established fact that a matrix has rank $k$ if it possesses $k$ nonzero singular values. When the data has lower intrinsic dimensionality, we can derive a simpler bound.

\vspace{0.25cm}
\begin{tcolorbox}[colback=GreenYellow]
\begin{corollary}\label{theory:partialSumCorollary}
	Suppose the data matrix has rank $k$ and assumptions of Theorem \ref{theory:partialSum} hold. Then the discrepancy between 
	$\|X\|_{2,k+1}^2$ and $(\|\bar{X}\|_{2,k}^2 + n \|\bar{x}\|^2)$ is absolutely bounded by the first squared singular value of a centered matrix $\bar{X}$, i.e., 
	\begin{equation*}
		| \|X\|_{2,k+1}^2 - 
		(\|\bar{X}\|_{2,k}^2 + n \|\bar{x}\|^2)| < \bar{\sigma}_1^2.
	\end{equation*}
\end{corollary}
\end{tcolorbox}
\vspace{0.25cm}

We conclude this section with the observation that the bound is not dependent on the true rank of a data matrix, as the right-hand side of the inequality is governed by $\bar{\sigma}_1$, irrespective of $k$, which results in a rather crude bound. It can also be postulated that the discrepancy is primarily explained by the projected variance onto the first principal component, independent of the true rank of the data matrix.

\section{Conclusion}\label{sec:conclusion}

In this paper, we revisited the classical problem of PCA from an algorithmic perspective. A common choice of implementation is to apply SVD onto the data matrix for efficient computation, which is a coveted property in an era where data is increasingly large and high-dimensional. While the method is straightforward, a critical point in the SVD-based approach is often overlooked: acquiring a basis for projection via SVD is only valid if the data matrix is centered. This detail is often bypassed by employing a folklore strategy to compute an extra base.

The main contribution of this paper is to study the discrepancy between true and approximate embeddings, the latter of which is obtained assuming the centering condition is violated. Our investigation offers theoretical characterizations of when the two results can be considered compatible in multiple aspects. We hope our analysis contributes to a better understanding of the core algorithmic principles for the SVD-based implementation of PCA among readers.

Looking forward, it could be fruitful to explore how the findings of this paper could impact the practice and teaching of PCA, particularly in relation to the use of the SVD. This could include developing new guidelines or resources to ensure that the critical step of mean centering is not overlooked. There may also be opportunities to develop improved or alternative heuristics that can offer the computational benefits of the SVD-based approach while ensuring alignment between embeddings. Further theoretical and empirical exploration of the circumstances under which the first right singular vector accurately accounts for the mean direction would be valuable. Finally, our investigation could serve as a foundation for studying other dimensionality reduction techniques and their algorithmic implementations, given the continued growth of high-dimensional data in science and other disciplines.

\bibliographystyle{dcu}
\bibliography{reference}

\pagebreak
\section*{Appendix}\label{sec:proof}

\subsubsection*{Proof of Proposition 1}

Since $\bar{X} = X - \bfone_n \bar{x}^\top$ for $\bar{x} = \sum_{i=1}^n x_i /n$, we have the following equality $\bar{X}^\top \bar{X} = X^\top X - n \bar{x} \bar{x}^\top$. It is well known that the largest eigenvector of $X^\top X$ is a maximizer of the Rayleigh-Ritz quotient,
\begin{equation}\label{eq:rrquotient}
	w^* = \underset{w \in \bbS^{p-1}}{\argmax}~ w^\top X^\top X w,
\end{equation}
for $\bbS^{p-1} = \lbrace x\in\bbR^p~|~\|x\|=1\rbrace$. Using the relationship between $X$ and $\bar{X}$, the quantity from Equation \eqref{eq:rrquotient} can be upper bounded as follows:
\begin{align*}
	\underset{w \in \bbS^{p-1}}{\max}~ w^\top X^\top Xw &= \underset{w \in \bbS^{p-1}}{\max}~w^\top \left(\bar{X}^\top \bar{X} + n \bar{x} \bar{x}^\top\right)w\\
	&\leq \underbrace{\underset{w \in \bbS^{p-1}}{\max}~ w^\top \bar{X}^\top \bar{X} \bar{x}}_{(\textrm{A})}
	 + n \cdot \underbrace{\underset{w \in \bbS^{p-1}}{\max}~ w^\top \bar{x}\bar{x}^\top w}_{(\textrm{B})}.
\end{align*}
According to the Cauchy-Schwarz theorem, (A) is maximized when $w = \bar{v}_1$ while (B) achieves its maximum when $w$ is parallel to $\bar{x}$. Therefore, the choice of $w=\bar{v}_1 = \bar{x}/\|\bar{x}\|$ attains the maximum as well as the equality by the Courant-Fischer theorem. \qed

\subsubsection*{Proof of Corollary 2}

Based on the assumption that $\bar{x}$ and $\bar{v}_1$ are parallel, we can rewrite the SVD of $X$ as follows:
\begin{align*}
	X &= \bar{X} + \bfone_n \bar{x}^\top\\
	&= \bar{X} + \bfone_n \|\bar{x}\| \bar{v}_1^\top \\
	&= \sum_{i=1}^p \bar{\sigma}_i \bar{u}_i \bar{v}_i^\top + \bfone_n \|\bar{x} \| \bar{v}_1^\top \\
	&= \sum_{i=2}^p \bar{\sigma}_i \bar{u}_i \bar{v}_i^\top + \left( \bar{\sigma}_1 \bar{u}_1 +  \bfone_n \|\bar{x} \|\right) \bar{v}_1^\top\\
	\intertext{Let $S = \bar{\sigma}_1 \bar{u}_1 + \bfone_n \|\bar{x}\|$ such that $\|S\|^2 = \bar{\sigma}_1^2 + n \|\bar{x}\|^2$, then the above equation continues as} 
	&= \sum_{i=2}^p \bar{\sigma}_i \bar{u}_i \bar{v}_i^\top + \|S\| \cdot \frac{S}{\|S\|} \bar{v}_1^\top \\
	&= \sum_{i=2}^p \bar{\sigma}_i \bar{u}_i \bar{v}_i^\top + \sqrt{\bar{\sigma }_1^2 + n\|\bar{x}\|^2} u^* \bar{v}_1^\top,
\end{align*}
where the first left singular vector of $X$ is 
\begin{equation*}
	u^* = \frac{S}{\|S\|} = \frac{\bar{\sigma}_1 \bar{u}_1 + \bfone_n \|\bar{x}\|}{\sqrt{\bar{\sigma }_1^2 + n\|\bar{x}\|^2}}. 
\end{equation*}\qed 

\subsubsection*{Proof of Proposition \ref{theory:shape1}}

It suffices to show the following inequality
\begin{equation*}
 \|Xz\|^2 =	z^\top X^\top Xz \leq z_0^\top X^\top X z_0 = \|Xz_0\|^2,
\end{equation*}
under the condition as the Courant-Fischer theorem completes the proof. Using the relationship
\begin{equation*}
	X^\top X = \bar{X}^\top \bar{X} + n\bar{x}\bar{x}^\top = \bar{V} \bar{\Sigma}^2 \bar{V}^\top + n\|\bar{x}\|^2 z_0 z_0^\top,
\end{equation*}
we first have 
\begin{align*}
	\|Xz\|^2 &= z^\top X^\top X z\\
	&= z^\top \left(\bar{V} \bar{\Sigma}^2 \bar{V}^\top + n\|\bar{x}\|^2 z_0 z_0^\top\right) z \\
&= \sum_{i=1}^p \bar{\sigma}_i^2 (z^\top \bar{v}_i)^2 + n\|\bar{x}\|^2 (z_0^\top z)^2.
\intertext{Similarly, we can derive the following}
\|Xz_0\|^2 &= z_0^\top X^\top X z_0 \\
&= \sum_{i=1}^p \bar{\sigma}_i^2 (z_0^\top \bar{v}_i)^2 + n \|\bar{x}\|^2 (z_0^\top z_0) \\
&=\sum_{i=1}^p \bar{\sigma}_i^2 (z_0^\top \bar{v}_i)^2 + n \|\bar{x}\|^2.
\end{align*}
Combining the above, the difference $\|Xz_0\|^2 - \|Xz\|^2$ is shown to be bounded below
\begin{align*}
	\|Xz_0\|^2 - \|Xz\|^2 &= n \|\bar{x}\|^2 + \sum_{i=1}^p \bar{\sigma}_i^2 (z_0^\top \bar{v}_i)^2 - n\|\bar{x}\|^2 (z_0^\top z)^2 - \sum_{i=1}^p \bar{\sigma}_i^2 (z^\top \bar{v}_i)^2 \\
	&= n \|\bar{x}\|^2 (1 - (z_0^\top z)^2) +\sum_{i=1}^p \bar{\sigma}_i^2 (z_0^\top \bar{v}_i)^2 - \sum_{i=1}^p \bar{\sigma}_i^2 (z^\top \bar{v}_i)^2\\
	&\geq n \|\bar{x}\|^2 (1 - (z_0^\top z)^2) + \bar{\sigma}_p^2 \sum_{i=1}^p (z_0^\top \bar{v}_i)^2 - \bar{\sigma}_1^2 \sum_{i=1}^p (z^\top \bar{v}_i)^2
	\intertext{and since $\lbrace \bar{v}_i \rbrace$ spans $\bbR^p$, by the Pythagorean theorem we have }
	&= n \|\bar{x}\|^2 (1 - (z_0^\top z)^2) + \bar{\sigma}_p^2 - \bar{\sigma}_1^2.
\end{align*}
Hence, if we bound the last line by 0,
\begin{equation*}
n \|\bar{x}\|^2 (1 - (z_0^\top z)^2) + \bar{\sigma}_p^2 - \bar{\sigma}_1^2 \geq 0
\iff \|\bar{x}\|^2 \geq \frac{\bar{\sigma}_1^2 - \bar{\sigma}_p^2}{n (1-(z_0^\top z)^2)},
\end{equation*}
we obtain the inscribed condition for which $z_0$ being the largest eigenvector of $X^\top X$.\qed

\subsubsection*{Proof of Corollary \ref{theory:shape2}}

It is easy to check that $z_0 \in \calU_{\epsilon}$ since $|z_0^\top z_0| = 1 > 1-\epsilon$ for any $\epsilon \in (0,1)$. Hence, it suffices to show that for any $z \notin \calU_{\epsilon}$, $\|Xz\|^2 \leq \|Xz_0\|^2$ if 
\begin{equation}\label{eq:proof_of_corollary}
	\|\bar{x}\|^2 \geq \frac{\bar{\sigma}_{\max}^2 - \bar{\sigma}_{\min}^2}{n\cdot \left(1 - (1-\epsilon)^2\right)}.
\end{equation}
This is because the statement $\|Xz\|^2 \leq \|Xz_0\|^2$ implies that all candidates for the first singular vector are contained in $\calU_{\epsilon}$. Fix $z \notin \calU_{\epsilon}$ such that $|z^\top z_0| < 1-\epsilon$, which yields 
\begin{eqnarray*}
\frac{\bar{\sigma}_{\max}^2 - \bar{\sigma}_{\min}^2}{1 - (z^\top z_0)^2}	 < \frac{\bar{\sigma}_{\max}^2 - \bar{\sigma}_{\min}^2}{1 - (1-\epsilon)^2}.
\end{eqnarray*}
By Proposition \ref{theory:shape1}, Equation \eqref{eq:proof_of_corollary} holds, which completes the proof. \qed

\subsubsection*{Proof of Proposition \ref{prop:3}}
Similar to before, first observe that the mean direction vector $z_0 = \bar{x}/\|\bar{x}\|$ can be written as a linear combination of right singular vectors, 
\begin{equation*}
	z_0 = \sum_{i=1}^p z_i \bar{v}_i = \bar{V} z,
\end{equation*}
for some $z_i \in \bbR,~i=1,\ldots,p$. Using the above, we can rewrite the unscaled covariance matrix of $X$ as follows:
\begin{equation}\label{eq:gram}
\begin{aligned}
	X^\top X &= \bar{X}^\top \bar{X} + n \bar{x} \bar{x}^\top \\
	&= \bar{V} \bar{\Sigma}^2 \bar{V}^\top + n \|\bar{x}\|^2 \bar{V} z z^\top \bar{V}^\top\\
	&= \bar{V} (
 \underbrace{\bar{\Sigma}^2 + n \|\bar{x}\|^2 zz^\top}_{(*)}	
	)\bar{V}^\top.
\end{aligned}
\end{equation}
It was shown from Corollary \ref{theory:corollary2} that the right singular vectors of $X$ and $\bar{X}$ are identical. Therefore, the last line of Equation \eqref{eq:gram} implies relationship between two sets of eigenvalues for $X^\top X$ and $\bar{X}^\top\bar{X}$. The quantity $(*)$ adds a diagonal matrix and an outer product matrix, which is called the diagonal-plus-rank-one (DPR1) matrix \citep{DPR1}. We employ an established result on spectrum for a given DPR1 matrix.

\begin{theorem}[Theorem 2.1 of \cite{Cuppen}]\label{interlacing}
	Let $D$ be a diagonal matrix, $D = \mathrm{diag} (d_1, \ldots, d_p), ~p\geq 2$ with $d_1 > d_2 > \cdots > d_p$ and $\mu\in\bbR^p$ a vector consisting of all non-zero elements. For a scalar $\rho > 0$, the eigenvalues of the matrix $D + \rho \mu\mu^\top$ are equal to the $p$ roots $\lambda_1 > \lambda_2 > \cdots  > \lambda_p$ of the rational function
	\begin{equation*}
		\begin{aligned}
			w(\lambda) &= 1 + \rho \mu^\top (D-\lambda I)^{-1} \mu \\
			&= 1 + \rho \sum_{j=1}^p \frac{\mu_j^2}{d_j - \lambda}.
		\end{aligned}
	\end{equation*}
	The corresponding eigenvectors $x_1, \ldots, x_p$ of $D+\rho \mu\mu^\top$ are given by 
	\begin{equation*}
		x_j = \frac{(D-\lambda_j I)^{-1}\mu}{\|(D-\lambda_j I)^{-1}\mu\|},
	\end{equation*}
	and the $d_j$ strictly separate the eigenvalues $\lambda_j$ as follows:
	\begin{equation*}
		d_p < \lambda_p < \cdots  < d_2 < \lambda_2 < d_1 < \lambda_1 < d_1 + \rho \mu^\top \mu.
	\end{equation*}
\end{theorem}

Denote $\lambda_j$ for the $j$-th squared singular value of $\bar{\Sigma}^2 + n \|\bar{x}\|^2 zz^\top$. Finally, plugging $D = \bar{\Sigma}^2,~\rho = n\|\bar{x}\|^2$ and $\mu = z$ into Theorem \ref{interlacing} completes the proof.\qed

\subsubsection*{Proof of Theorem \ref{theory:partialSum}}
We use the same reasoning to represent the mean direction $z_0$ as an element in the space of $\bar{V}$, i.e., $z_0 = \bar{V}z$. The first singular value of an unscaled matrix $X$ can be written as 
\begin{equation*}
	\begin{aligned}
		\|Xz_0\|^2 &= z_0 X^\top X z_0\\
		&= z_0^\top \bar{V} ( \bar{\Sigma}^2 + n\|\bar{x}\|^2 zz^\top )\bar{V}^\top z_0\\
		&= z^\top \bar{V}^\top \bar{V} ( \bar{\Sigma}^2 + n\|\bar{x}\|^2 zz^\top )\bar{V}^\top \bar{V} z\\
		&= z^\top \bar{\Sigma}^2 z + n \|\bar{x}\|^2 (z^\top z)^2\\
		&= z^\top \bar{\Sigma}^2 z + n \|\bar{x}\|^2,
	\end{aligned}
\end{equation*}
using the relationship revealed from Equation \eqref{eq:gram} and the facts that $\bar{V}^\top \bar{V} = I$ and $z^\top z = 1$. From Proposition \ref{prop:3} and the above, we get a bound for $\sigma_1^2$ as
\begin{eqnarray}\label{eq:maxBound}
	\max\lbrace\bar{\sigma}_1^2, \bar{\sigma}_p^2 + n\|\bar{x}\|^2\rbrace =  \bar{\sigma}_p^2 + n\|\bar{x}\|^2 \leq \sigma_1^2 \leq \bar{\sigma}_1^2 + n\|\bar{x}\|^2.
\end{eqnarray}
The maximum value from Equation \eqref{eq:maxBound} is obtained from an assumption that the norm of the mean vector is large enough as 
\begin{equation*}
	\|\bar{x}\|^2 \geq \frac{1}{n} \frac{\bar{\sigma}_1^2 - \bar{\sigma}_p^2}{1 - (1-\epsilon)^2} \geq \frac{\bar{\sigma}_1^2 - \bar{\sigma}_p^2}{n}.
\end{equation*}
This leads to compare the partial sums of squared singular values of $X$ and $\bar{X}$ as 
\begin{equation*}
\sum_{j=2}^{k+1} \bar{\sigma}_j^2 + n \|\bar{x}\|^2 < \sum_{j=1}^{k+1} \sigma_j^2 < 2 \bar{\sigma}_1^2 + \sum_{j=2}^k \bar{\sigma}_j^2 + n \|\bar{x}\|^2. 
\end{equation*}
Therefore, the difference of $\|X\|_{2,k+1}^2$ and $\|\bar{X}\|_{2,k}^2 + n \|\bar{x}\|^2$ is bounded as follows;
\begin{equation*}
- \bar{\sigma}_1^2 + \bar{\sigma}_{k+1}^2 < \|X\|_{2,k+1}^2 - (\|\bar{X}\|_{2,k}^2 + n \|\bar{x}\|^2) < \bar{\sigma}_1^2.
\end{equation*}\qed

\subsubsection*{Proof of Corollary \ref{theory:partialSumCorollary}}
It is stated that the data matrix $X$ is rank $k$. This is equivalent to state that the centered version $\bar{X}$ is also of rank $k$ so that $\bar{\sigma}_j = 0$ for all $j > k$. Therefore, the inequality from Theorem \ref{theory:partialSum} can be written as 
\begin{equation*}
	- \bar{\sigma}_1^2 < \|X\|_{2,k+1}^2 - 
	(\|\bar{X}\|_{2,k}^2 + n \|\bar{x}\|^2) < \bar{\sigma}_1^2,
\end{equation*}
which completes the proof. \qed

\end{document}